\documentstyle[preprint,aps,floats]{revtex}
\tightenlines

\input psfig.tex 

\begin{document}

\def\ba{\begin{eqnarray}}
\def\ea{\end{eqnarray}}
\def\be{\begin{equation}}
\def\ee{\end{equation}}
\def\({\left(}
\def\){\right)}
\def\[{\left[}
\def\]{\right]}
\def\lagrange {{\cal L}}
\def\del {\nabla}
\def\d {\partial}
\def\Tr{{\rm Tr}}
\def\half{{1\over 2}}
\def\fourth{{1\over 8}}
\def\bibi{\bibitem}
\def\S{{\cal S}}
\def\H{{\cal H}}
\def\xx{\mbox{\boldmath $x$}}
\newcommand{\phpr} {\phi_0^{\prime}}
\newcommand{\gam}{\gamma_{ij}}
\newcommand{\sqgam}{\sqrt{\gamma}}
\newcommand{\dph}{\delta\phi}
\newcommand{\om} {\Omega}
\newcommand{\dom}{\delta^{(3)}\left(\Omega\right)}
\newcommand{\rar}{\rightarrow}
\newcommand{\Rar}{\Rightarrow}
\newcommand{\labeq}[1] {\label{eq:#1}}
\newcommand{\eqn}[1] {(\ref{eq:#1})}
\newcommand{\labfig}[1] {\label{fig:#1}}
\newcommand{\fig}[1] {\ref{fig:#1}}
\def\gsim{ \lower .75ex \hbox{$\sim$} \llap{\raise .27ex \hbox{$>$}} }
\def\lsim{ \lower .75ex \hbox{$\sim$} \llap{\raise .27ex \hbox{$<$}} }
\newcommand\bigdot[1] {\stackrel{\mbox{{\huge .}}}{#1}}
\newcommand\bigddot[1] {\stackrel{\mbox{{\huge ..}}}{#1}}


\def\dm{\frac{m'}{m}}
\def\dn{\frac{n'}{n}}
\def\ddm{\frac{m''}{m}}
\def\ddn{\frac{n''}{n}}
\def\nl{\nonumber \\}

\title{\bf Singular Cosmological Instantons Made Regular}
\author{
Kelley Kirklin 
\thanks{Email:K.H.Kirklin@damtp.cam.ac.uk}, Neil
Turok\thanks{Email:N.G.Turok@damtp.cam.ac.uk}
and Toby Wiseman \thanks{Email:T.A.J.Wiseman@damtp.cam.ac.uk}}
\address{
 DAMTP, Centre for Mathematical Sciences, Cambridge, CB3 0WA, U.K.}
\date{\today}
\maketitle

\begin{abstract}
The singularity present in cosmological instantons of
the Hawking-Turok type is resolved by a conformal transformation,
where the conformal factor has a linear zero of codimension one.
We show that if the underlying regular manifold is taken to 
have the topology of $RP^4$, and the conformal factor is taken
to be a twisted field so that the zero is enforced, then one
obtains a one-parameter family of solutions of the classical
field equations, where the minimal action solution has 
the conformal zero located on a minimal volume
noncontractible $RP^3$ submanifold. 
For instantons with two singularities, the corresponding topology 
is that of a cylinder $S^3\times [0,1]$ with $D=4$ analogues of 
`cross-caps' 
at each of the endpoints.

\end{abstract}
\vskip .2in

\section{Introduction}

Euclidean quantum gravity provides an approach to 
some of the most fundamental issues in cosmology. 
One of these
is the question of the initial state of the Universe,
both for the background geometry and for the fluctuations. 
Euclidean methods have long been applied to calculations of
quantum fluctuations in inflation, 
and to tunnelling  problems in
de Sitter space. 

A recent development was the observation that a generic theory of
scalar matter coupled to gravity allows a one-parameter family
of singular but finite action Euclidean instantons 
which can be used to describe the beginning of 
inflating open or closed universes\cite{ht},\cite{bl}. 
The free parameter in these solutions is just the value of the 
cosmological density parameter $\Omega$ today.

The singular nature of these instantons is a cause for concern since
it is not clear whether the classical field equations are satisfied at
the singularity.
In references 
\cite{ntplb}  it was argued that the solutions 
should be regarded as constrained instantons, described by a collective 
coordinate to be integrated over in the path integral.

In this paper we clarify what the collective coordinate is 
and how it is to be integrated over. The singularity is resolved by a 
conformal transformation. The original `Einstein frame'
 metric equals a regular underlying metric
times a conformal factor with a linear zero of codimension one.
The regular metric describes an  $RP^4$ and the conformal factor is
taken to be a section of the nontrivial $Z_2$ 
orientation bundle over $RP^4$. Because of the nontrivial 
twist, the action is not guaranteed to be stationary for 
solutions of
the classical differential equations of motion: additional 
data enters on a nontrivial three manifold upon which 
the conformal factor vanishes. We draw an analogy with
magnetic monopole solutions on $S^2$, where one 
also has a one parameter family of solutions to the
classical field equations of varying action. 
In our case, when we stationarise the action with 
respect to the free parameter, we find a lowest action 
classical solution.
We extend these considerations
to instantons representing the beginning of closed inflationary universes,
with topology corresponding to $S^3\times [0,1]$ with cross
caps at either end. 

Allowing the conformal factor to vanish 
is clearly incompatible 
with having a 
a globally Riemannian manifold with positive definite metric.
However, there are reasons for believing that in a quantum theory
of geometry such behaviour is inevitable. Our best examples of 
such geometrical theories are string theory, and two plus one dimensional
gravity. In the former, if one takes the Lorentzian path integral 
seriously, it is not possible to have a globally Lorentzian 
metric on worldsheets with genus not equal to one. There must be
singular points at which the determinant of the world-sheet 
metric vanishes. Likewise, in two plus one dimensional gravity, 
it has long been argued that one should also take into account
vierbeins which have vanishing determinant.

We believe that the interpretation given here 
resolves some other worries 
which have been expressed regarding singular instantons. Since 
the singularity introduces a `conformal boundary', besides apparently
violating the intent of the `no boundary' proposal, 
conformally coupled radiation might be able to enter or leave the 
spacetime in an arbitrary manner. In our interpretation, where
the apparent `conformal boundary' has antipodal points identified,
the underlying smooth manifold is compact and there is no boundary.
Likewise the concern raised by Vilenkin \cite{Vil} that
a `necklace' of constrained instantons would have 
arbitrarily 
negative Euclidean 
action is also resolved because the number of 
surfaces on which the constraint enters is determined topologically.
For solutions of maximal $O(4)$ symmetry, the number of such surfaces 
can only be 0, 1 or 2 and we shall discuss the last two cases
here. In our construction, 
`necklaces' do not occur as solutions of the classical
field equations.

\section{Review}

The singular instantons described in \cite{ht} are O(4) invariant 
solutions with line element 
\be
ds^2= d\sigma^2 +b^2(\sigma)d \Omega_3^2
\labeq{emetric}
\ee
where $d \Omega_3^2$ is the round three sphere metric. 
The Euclidean field equations governing the metric and scalar field $\phi$
are 
\be 
b_{,\sigma \sigma}=- \kappa {b\over 3} \left(\phi_{,\sigma}^2 +V(\phi)\right),
\qquad b_{,\sigma}^2 = \kappa {b^2\over 3}\left({1\over 2} \phi_{,\sigma}^2 -V(\phi)\right) +1,
\qquad (\phi_{,\sigma}b^3)_{,\sigma} = b^3 V_{,\phi}(\phi).
\labeq{fes}
\ee
Here and below we set 
$\kappa= 8\pi G$ where $G$ is Newton's constant. 
As long as the potential $V(\phi)$ is not too steep at large 
$\phi$ 
there is a one parameter
family of finite action 
solutions in which the scalar field starts at some
$\phi_0$ and then rolls uphill. The scale factor $b(\sigma)\sim
 \sigma $
at the regular pole of the instanton, where $\sigma=0$ and 
$\phi=\phi_0$. As 
$\sigma$ increases, $b(\sigma)$ takes the form of a deformed 
sine function. As $b$ approaches its second zero, 
the scalar field's motion is antidamped and it runs off to infinity.
At the singularity $\sigma_m$, the scale factor $b$  vanishes as
 $(\sigma_m-\sigma)^{1\over 3}$,
and  $\phi$ diverges logarithmically,
$\phi \approx - \sqrt{2\over 3\kappa } {\rm ln}\left(
C(\sigma_m-\sigma)\right)$. From this behaviour it
follows that $b^2 e^{\sqrt{2\kappa \over 3}\phi}$ tends to a constant
at the singularity. This constant will turn out to be the
`radius squared' of the zero conformal factor locus in the 
regular underlying metric, and will play the role of the
collective coordinate mentioned in the Introduction.

The divergence of $\phi$, and of the Ricci scalar for the metric
at the singular point $\sigma_m$, may seem physically unreasonable,
but the finiteness of the action tells us that we should take these
singularities seriously since they are not obviously 
suppressed in the path integral. In fact we shall show that by
a suitable change of variables on superspace, the singularity
may be removed thus making the action finite term by term.
Another possible complaint is that 
we have no reason to suppose simple behaviour for the 
potential $V(\phi)$ at field values much greater than the 
Planck mass. But whilst
the singular instanton solutions 
do probe arbitrarily large values of $\phi$, calculations
of observable quantities such as the density
perturbations are very insensitive to the precise form
of the potential at large $\phi$, precisely because the potential 
itself (provided it is not very steep) plays very little role 
in the vicinity of the
singularity. In any case, the theory applies virtually
unchanged to potentials which are bounded above and
therefore never produced super-Planckian energy densities. 

A clue to the interpretation of singular instantons
is obtained by rewriting the metric in the 
form $b^2(X) (dX^2 +d \Omega_3^2)$. One sets $dX = d \sigma /b$, thus
$X\propto \sigma^{2\over 3}$. From this it follows that as one approaches
the singularity, $b^2(X)$
vanishes linearly with $X$, so
that the singularity
is a {\it linear zero of the conformal factor,
of codimension one}. Solutions with singularities of the same
character were 
discovered in supergravity some time ago. They describe
two dimensional
`tear-drop compactifications' of ten dimensional supergravity
\cite{gz}. As noted by those authors, although the
relevant manifolds are noncompact, they possess
many desirable properties, including a quantised 
mass spectrum and unbroken supersymmetry to protect 
against quantum fluctuations. (Incidentally they also have a 
chiral spectrum of zero modes, and were in some respects the
antecedents of the now more popular orbifold compactifications
of eleven dimensional supergravity).

The simple nature of the singularity suggests the
interpretation we shall explore 
below, namely that
the conformal factor is a field forced to vanish by a
topological constraint. We discuss a suitable constraint 
in the next section.

\section{Twisted fields}

It is a familiar notion that in infinite space, 
field theories with degenerate vacua possess topologically stable
soliton solutions. The 
condition of finite
energy forces the fields to lie in vacuo at infinity. If
the map defined by the fields at infinity onto 
the vacuum manifold is topologically nontrivial, the field is forced
to vanish at isolated points, and solitons occur at these points. 
Solitons like 
these are in general only strictly stable if space is infinite. 
But on 
finite spaces it is still possible to have zeros
enforced topologically, and therefore have topologically stable solitons. 
This occurs if the field configuration is `twisted'. This option 
exists if there are noncontractible loops on the manifold, and if
fields can aquire a minus sign as these loops are traversed.  
In mathematical terminology a twisted field 
is a
section of a nontrivial fibre bundle, requiring more than one coordinate
chart for its definition. 
The simplest case is
a scalar field theory on a circle with a
$Z_2$ internal symmetry $\phi \rightarrow -\phi$. We 
have two choices of boundary conditions for $\phi$
 - periodic or antiperiodic, see Figure (\ref{fig:circle}).
Both are equally natural because there is no
physical distinction between 
$+\phi$ and $-\phi$.
As the coordinate increases by the length $L$ of the circle, there
is no reason to match $\phi$ to $+\phi$ rather than $-\phi$.
In the first case,
the configuration space is a trivial bundle over $S^1$, but in
the second it is a nontrivial bundle, and the scalar field
aquires 
a $-1$ as one passes through the single nontrivial coordinate transition.
In the path integral there is no reason not to sum
over both the twisted and untwisted sectors.

\begin{figure}
\centerline{\psfig{file=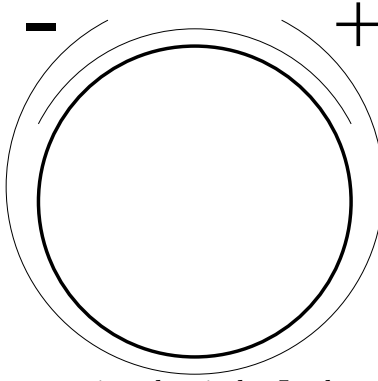,width=2.in}}

\caption{Two coordinate patches covering the circle. In the twisted
field sector one must reverse the sign of the scalar field when passing 
between the two coordinate regions on one
of the two overlaps (e.g. the one labelled $-$), but not on the other
(labelled $+$). 
}
\labfig{circle}
\end{figure}
In the twisted sector of a $Z_2$ symmetric scalar field theory, the field
must vanish somewhere. The interesting case is where the scalar potential
yields spontaneous 
symmetry breaking, for example $V=-m^2 \phi^2 +\lambda \phi^4$. In this case, 
at least for large $L$, 
energetic considerations prefer that the 
the field be nonzero over most of space. In the twisted 
sector one must have an odd
number of zeros of $\phi$, 
whereas in the untwisted sector there must be an even number.
For finite $L$ the two vacua involved 
in each case will mix quantum mechanically, with the symmetric state
being the ground state. 
The theory therefore splits into sectors
labelled by a $Z_2$ topological charge, equal to $(-1)^N$ where 
$N$ is the number of zeros. 

To define the action for twisted fields, one must add
the contributions from each coordinate patch. 
Only one of the two transitions between coordinate patches is nontrivial,
so one may take the action to be a single integral evaluated in 
a single patch, running from $0$ to $L$. The only problem is that 
we have to differentiate the field across the special point
$x=0$, identified
with $x=L$. The twisted field must undergo a sign change as one crosses 
this point. The way to differentiate is to note that within a single
coordinate patch the derivative is defined as usual as
$d\phi/dx= $ Lim$\epsilon \downarrow 0$ 
$(\phi(x+\epsilon/2)-\phi(x-\epsilon/2))/\epsilon$. But if one 
differentiates across the singular point, one must include a 
compensating minus sign, using instead
Lim$\epsilon \downarrow 0$  
$(\phi(x+\epsilon/2)+\phi(x-\epsilon/2))/\epsilon$. 
We define the latter as the covariant derivative of the field,
$D\phi$. 
(One could
instead insert the minus sign in front of $\phi(x+\epsilon/2)$,
which would reverse the sign of $D\phi$. But like $\phi$ itself,
$D\phi$ is only defined up to a sign and nothing physical changes 
if one reverses it.)

The action is then given by
\be
{\cal S}= \int dx \left[{1\over 2} ( D
\phi )^2 +V(\phi)\right].
\labeq{acttw}
\ee

\begin{figure}
\centerline{\psfig{file=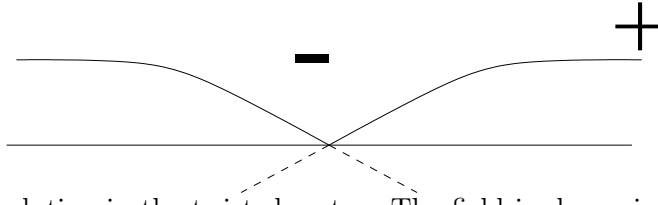,width=3.5in}}
\caption{A soliton solution in the twisted sector. The field is
shown in each of two coordinate patches. There is a trivial 
coordinate overlap (+) where the field is identified at $0$ and $L$,
and a nontrivial overlap (-) where the field reverses sign. The
solution shown is continuous: this requires that one place the
transition between the coordinate patches 
at the zero of the field. However the transition may be placed anywhere,
for example at the maximal value of the field.
For any location the 
action is stationary as long as
the covariant derivative is continuous (see text).
}
\labfig{twisted}
\end{figure}

For example, choosing the singular point to be located at $x_0=0$ the action 
is an integral running from $0$ to $L$. It is varied
subject to the boundary constraint that $\phi(0)=\pm \phi(L)$
in the untwisted and twisted cases, obtaining
\be
{\delta {\cal S}} = \left[\delta \phi {d\phi \over dx} \right]_0^L +
\int_0^L dx \delta \phi\left[-{d^2 \phi \over d x^2}+V_{,\phi}\right].
\labeq{acttv}
\ee
The action is stationary under all field variations about 
a classical solution $\phi_c(x)$ if 
$\phi_c(x)$ obeys the classical field equations away from $x=0$ and
if ${d \over dx } \phi (0)=  \pm {d \over dx }
\phi (L)$, which is just the requirement that the
covariant derivative be continuous. 

For each value of $L$ there is a stationary configuration in the 
twisted sector. For small $L$ this configuration is just $\phi=0$. This 
configuration is common to both the twisted and untwisted sectors,
but it is only stable in the twisted sector
because the destabilising 
negative mode $\delta \phi =$ constant
is disallowed by the twisted boundary conditions. 

For larger $L> \pi m^{-1}$ one can easily check that 
$\phi=0$ is unstable to a twisted negative mode, and the lowest action
solution spontaneously breaks the
$Z_2$, taking the form shown in
Figure 2, with a single zero of the scalar field. We shall show that
this solution provides a
stable twisted classical vacuum. 
In the gauge where $\phi$  is continuous there is a
`kink' $d \phi/dx|^+_-$ at the
zero of $\phi$, which looks like a charged source coupling to
$\phi$. But this `charge' is just a reflection
of the change of coordinate chart across the zero.

To find the minimum action state
it is convenient to label each field configuration by the
by the maximal value of $|\phi|$. Without loss of generality
we can take this value of $\phi=\phi_m$
to  be positive.
We can also take the sign flip 
of the field to occur there. Now the field runs from 
$-\phi_m$ to $+\phi_m$, as $x$ runs from $0$ to $L$.
However note that we cannot (except by
working on a larger covering space - see below) construct
a globally valid action involving ordinary integrals 
and derivatives of the fields. We have to proceed
by first introducing an `internal boundary' with 
the field taking the value $\phi_m$, and then 
treat $\phi_m$ as a constrained parameter when
we vary the action to obtain the field equations.

We start by assuming the field is monotonic.
Now we write the action integral as
\be
\int_0^L dx {1\over 2} \left( 
\phi' - \sqrt{ 2 (V(\phi)-V_{m})}\right)^2 +2 
\int_0^{\phi_{m}} d \phi\sqrt{2\left( V(\phi)-V_{m}\right)}+L
V_m,
\labeq{acttwa}
\ee
where $V_m\equiv V(\phi_m)$.
The
first term is positive semidefinite.
The action is then bounded below by the value of
the second two terms, which we may minimise with
respect to $\phi_m$.
For any 
$\phi_m$,
the first term is minimised by an appropriate solution
of the classical field equations. For the $\phi_m$ which
minimises the second two terms, the minimum of the first term
is zero.
Any other field configuration clearly has larger action and 
therefore the classical solution is
absolutely stable.

If we minimise the second two terms we find
\be
V_{,\phi}(\phi_m)\left(
2 \int_0^{\phi_{m}} {d \phi \over \sqrt{ 2 (V(\phi)-V_{m})}}-L\right)=0.
\labeq{phieqb}
\ee
The nontrivial solution is that which occurs when 
$V_{,\phi}(\phi_m)$ is not zero. If this solution exists,
we can show it is a minimum by changing variables in
the integral 
 to  $y=\phi/\phi_m$ which 
runs from $0$ to $1$. After absorbing the $\phi_m$ in the
denominator, the only 
$\phi_m$ dependence is in the quartic (or more generally the
non-quadratic 
terms) of $V$ and $V_m$. Hence we see that the first
term in (\ref{eq:phieqb}) is monotonically increasing with $\phi_m$.
This, with the fact that $V_{,\phi}(\phi_m)$ is negative,
proves that the second two terms in (\ref{eq:acttwa}) increase
away from the stationary point.

The minimum of the 
the first term in (\ref{eq:acttwa}) is obtained when the Bogomol'ny
equation
$\phi' = \sqrt{ 2 (V(\phi)-V_{m})}$ holds.
By integrating this we obtain a relation
between $L$ and $\phi_m$. But this is precisely the condition that
the bracketed term in (\ref{eq:phieqb}) vanishes.
Thus the solution to the Bogomol'ny equation
minimises the
action.

We assumed monotonicity above but it is not hard to show that
the energy is always greater for non-monotonic configurations.
Notice that stability depends on the
higher power terms in the potential - a purely quadratic potential
does not allow any nontrivial classical solution except $\phi=0$.
In the example we have chosen one can perform the
integrals as elliptic integrals K and exhibit the `critical behaviour'
in $\phi_m$ for $L$ just above $\pi m^{-1}$,
 $\phi_m \propto \lambda^{-{1\over 2}} (L-\pi m^{-1})^{1\over 2}$,
with action proportional to
$ - \lambda 
\phi_{m}^4$. 

Note that it was useful in this analyis to
label field configurations
by the maximal field value $\phi_m$, separately minimising the
action with respect to $\phi_m$ on the `internal boundary' 
and with respect to
variations away from the boundary.
The Bogomol'ny equation for example explicitly depends 
on $\phi_m$. 
This is not a procedure one is used to for untwisted fields,
because the ground state configuration is
trivial and independent of $L$. In contrast the
twisted vacuum 
depends strongly on $L$, even exhibiting a `phase 
transition' at $L= \pi m^{-1}$. 

In the gravitational instanton case we shall adopt
a similar strategy, with $\phi_m$ 
replaced by the maximal value of a certain field $n$ 
which shall be twisted.
The size of the instanton, analogous to $L$, 
will also be integrated over. Hence we find
a one parameter family of classical solutions 
for each value of 
either the maximal field value or alternatively the size
of the instanton.
The action is minimised
by one particular solution of the classical
field equations.

For a twisted scalar field on a circle, we could have 
represented the problem on the covering space, 
a circle twice as large, where we used only the odd Fourier
modes for the twisted field. In the nonorientable four dimensional
example below this option is not available to us, since
we wish to consider
an action density which is odd under the $Z_2$. 
If we integrated naively on the covering space we
would obtain zero. Instead we must include an orientation flip factor
in the integral, which effectively reduces it to one over
half the covering space.

\section{$RP^4$}

The $Z_2$ we discussed in the previous section was a purely 
internal symmetry. Next we shall identify a similar $Z_2$ symmetry 
acting on the conformal factor and related to orientation reversals
for coordinate patches covering a non-orientable manifold.
We are interested in viewing the 
conformal factor as a twisted field. We therefore consider 
metrics of the form 
\be
g_{\mu \nu} (x)= \Omega^2 (x) g_{\mu \nu}^{R} (x)
\ee
where $g_{\mu \nu}^{R}(x)$ is a Riemannian (positive definite)
metric which shall be regular in the classical solution
but the conformal factor
$\Omega^2(x)$ is allowed to go negative. We shall in what follows
refer to geometrical quantities calculated in the metric 
$g_{\mu \nu}^{R} $ as being in the `Riemannian frame', using
terminology analogous to that used in string theory where
one considers the `string frame' or `Einstein frame', which are
related by a conformal transformation involving the dilaton field.

We consider a 
theory with a local $Z_2$ symmetry, where the $Z_2$ involves changing
orientation and, in the twisted sector of the theory, 
reversing the sign of $\Omega^2$. With such a local symmetry
we can always choose $\Omega^2$ to be positive everywhere
except at conformal zeros. 
Therefore we
are not considering `anti-Euclidean' or `mixed signature'
spacetimes, but we are
allowing zeros of the conformal factor to be topologically 
enforced. \footnote{These
conformal zeros will be wrapped around noncontractible
codimension one submanifolds of the of the Euclidian spacetime, and may
therefore be viewed as stable domain walls. There is a natural generalisation
of this construction to higher codimensions. For example, allowing $\Omega$ to
carry $U(1)$ rather than $Z_{2}$ charge will lead to stable codimension
two string worldsheet conformal zeros wrapped on noncontractible two cycles in
spacetime. This rich class of structures is under investigation.\cite{ktprep}
}

Let us now explain the reason for linking the sign change of 
$\Omega^2$ with orientation reversal. 
The gravitational/scalar action density (discussed below) 
involves terms linear in 
$\Omega^2$ and is therefore odd under the $Z_2$. 
The only way to compensate for a sign change $\Omega^2 \rightarrow 
-\Omega^2$ 
is to have the 
the integration measure $d^4 x$ change sign
under the same $Z_2$. This means that for the action to
be invariant the orientation of the 
coordinate system 
must change each time $\Omega^2$ changes sign. 
Twisted fields by definition undergo 
an odd number of sign changes as one circumnavigates the
background space along certain noncontractible paths. 
If $d^4 x$ is to do the same, the
manifold must be non-orientable, and we must identify the 
$Z_2$ of the twisted line bundle 
with the $Z_2$ of the orientation
bundle. 

$RP^4$ is the obvious candidate manifold, obtained from
$S^4$ by identifying antipodal points. There 
is no global choice for orientation on it.
The orientation bundle over $RP^4$ has a $Z_2$ structure group, and
we shall identify some scalar fields (including $\Omega^2$) 
as odd and twisted under this $Z_2$. We shall 
obtain an invariant action in the twisted sector, and the
singular
instantons of Section II will emerge as solutions of 
the classical field equations in this
sector.

\begin{figure}
\centerline{\psfig{file=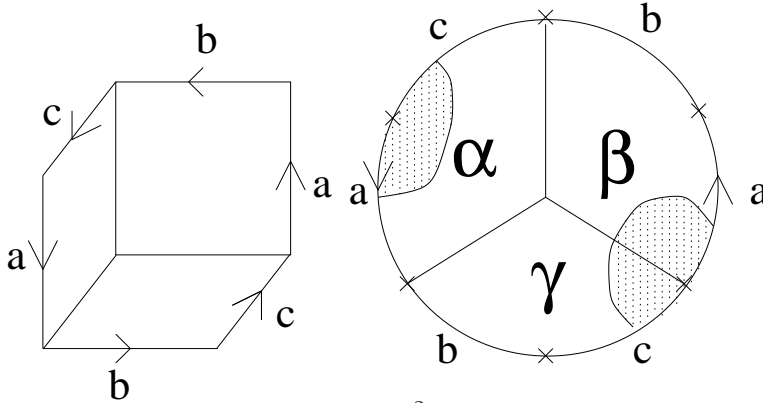,width=4.in}}
\caption{Three coordinate patches covering $RP^2$, forming
three of the six faces of a cube and with the identifications indicated by
arrows. The boundary of the three patches is an $RP^1$ which
plays the role of an `internal boundary ' when one constructs the action
integral. 
}
\labfig{rp2}
\end{figure}

The space $RP^2$ is easier to 
visualise than $RP^4$ and shares the features of 
non-orientability, and a class of 
noncontractible codimension one submanifolds, which shall
be central to the discussion below. 
$RP^2$ is illustrated in Figure (\ref{fig:rp2}). 
A sphere may be projectively mapped in a one to 
one manner onto 
the surface of a cube.
$RP^2$ is then formed from 
three faces of the 
cube by identifying edges as shown in the diagram 
on the left in Figure (\ref{fig:rp2}). We can further simplify the
diagram by mapping the whole onto a disk as on the right of
Figure (\ref{fig:rp2}). The 
coordinate patches may be extended into small overlap regions.
Each of the three patches
connects to the other two via two alternate overlap regions.
The first involves crossing the labelled boundary $abc$ 
and a change in orientation, 
Jacobian $det(\partial x'/\partial x)$ being negative. 
The other does not, and the Jacobian is positive. 
(This is made clear by using the obvious
Cartesian coordinates on the three faces of the cube).
The second important feature is that there is a class of 
noncontractible $RP^1$'s on $RP^2$, the most obvious member
of which is the $RP^1$ boundary $abc$ of the three
faces shown in Figure (\ref{fig:rp2}) with appropriate
identifications. (Note that $RP^1$ is isomorphic to $S^1$: 
but the same is not true of $RP^n$ for $n>1$). 
More generally 
any closed path 
in $RP^2$ which intersects the $RP^1$ an odd number
of times is noncontractible, and as one travels along it one 
must pass through an odd number of nontrivial coordinate transitions. 

An important check that we have a consistent fibre bundle is
that the product of $Z_2$ group elements in triple
overlap regions, $g_{\alpha \beta} g_{\beta \gamma} g_{\gamma \alpha}$
should be unity. This is indeed satisfied here because in such
a triple overlap (shaded region on the right of Figure 
(\ref{fig:rp2}) two of the transitions involve 
a charge of orientation and the other does not.
In the case of $RP^4$ the construction is completely
analogous except we take half of the faces 
of a five dimensional
hypercube. There is a class of noncontractible $RP^3$'s
analogous to the $RP^1$'s here and as
for $RP^2$ one can pick one of them to be the one across which
the nontrivial field sign and orientation flips occur.

The action for twisted fields takes the form
\be
S= \Sigma_i \int d^4x_i \sqrt{g} L_i
\labeq{acti}
\ee
where the sum runs over coordinate patches, and the integral is
broken into non-overlapping pieces with common 
boundaries within
the coordinate overlaps. If the Lagrangian density is a twisted scalar,
then moving these boundaries around the manifold
does not change the action since the minus sign from the Jacobian 
$det(\partial x /\partial x')$
is compensated for by the minus sign aquired by the Lagrangian density.
Note that the determinant $g$, according to the usual
transformation laws does {\it not} aquire a sign change under
an orientation flip and so it is an `untwisted' tensor density.

The action is a functional of the fields in
the `bulk' of the $RP^4$, with an `internal boundary' which is $RP^3$.
To write it explicitly we may employ the covering space
$S^4$, which consists of two identical copies of the $RP^4$.
Twisted fields are then just
odd parity functions on $S^4$ while untwisted fields are
even parity functions. The two copies of $RP^4$ are joined on
an even parity three-surface, which consists of two copies of $RP^3$. 
However there is a subtlety associated with integration over 
the $RP^4$.
As we have explained, the integral of a twisted scalar is perfectly well
defined on $RP^4$. But if we use the naive integration measure
on $S^4$, a twisted scalar would integrate to zero. Instead
we must define the integration measure by
building  in an  orientation flip on the three surface common to
the 
$RP^4$'s. This involves multiplying 
$d^4x \sqrt{g}$ by a twisted function $\epsilon(x)$ which
equals 
$+1$ on one side and $\epsilon=-1$ on the other side
(Figure (\ref{fig:coverspace})
Now when we integrate a twisted (odd parity) field
over the $S^4$ and divide by two we get the 
correct integral over $RP^4$. 
The function $\epsilon(x)$ effectively introduces a boundary 
into the problem, which is as we have explained an 
$RP^3$ since twisted or untwisted fields must be
odd or even parity on it as well. This `internal boundary'
we introduce has degrees of freedom associated with 
its location on the $S^4$. It is not the boundary of the manifold
however, rather it is the location of the orientation flip 
which occurs as we circumnavigate $RP^4$. 

\begin{figure}
\centerline{\psfig{file=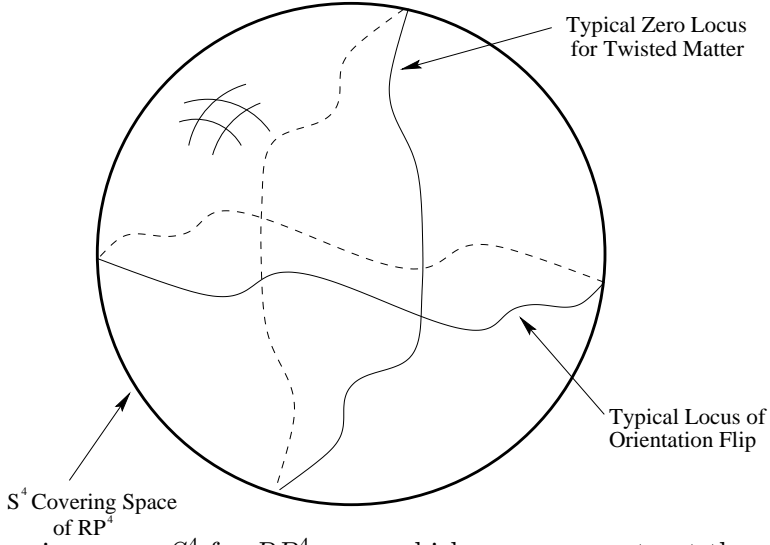,width=4.in}}
\caption{The covering space $S^4$ for $RP^4$ upon which we may 
construct the action. The twisted fields are odd parity on $S^4$ and therefore
must possess codimension one zeros. When we integrate a twisted field
on $RP^4$ we must use the measure $\int d^4 x \epsilon(x)$ on $S^4$ where
$\epsilon(x)$ is an odd parity field equal to $+1$ on half of the 
$S^4$. The three manifold upon which $\epsilon$ changes sign is 
noncontractible on $RP^4$, and plays the role of a boundary when constructing
the action. 
}
\labfig{coverspace}
\end{figure}

\section{The Action for Twisted Fields}

We are now ready to consider the Einstein-scalar theory discussed in
Section II. Our first task is to remove the singularity in the metric to
obtain well defined field equations. This is done by
changing
coordinates on superspace to fields which are regular everywhere. 
The procedure is familiar in the case
of singular spacetime coordinates, for example the 
usual Schwarzchild coordinates for a black hole. 
Here, as there, the purpose of the change in 
coordinates is to enable 
us to pass through the singularity in an unambiguous way
to see what is on the other side. 

As mentioned above we consider 
metrics
of the form
\be
g_{\mu \nu} (x)= \Omega^2 (x) g_{\mu \nu}^{R} (x)
\labeq{sec}
\ee
where the `underlying' metric $g_{\mu \nu}^{R}$ on a compact four-manifold 
${\cal M}$ is
assumed positive definite and the 
conformal factor $\Omega^2 (x) $ is viewed 
as a scalar field living on ${\cal M}$ which
is allowed to vanish. The metric $g^R$ shall be regular in the
classical solutions we discuss. 
Writing the metric this way introduces an
obvious local 
 symmetry in our (redundant) description of the theory, namely the conformal 
symmetry
\be
\Omega^2(x) \rightarrow
\omega^{-2}(x) \Omega^2(x), \qquad g_{\mu \nu}^R(x)\rightarrow \omega^2(x)
g_{\mu \nu}^R(x),
\labeq{conf}
\ee
and we shall adopt this symmetry as
fundamental in the construction below. 

When we write the action integral for $RP^4$ we
need to take into account the `internal boundary' mentioned above.
We can write the integral over $S^4$ with an extra function 
$\epsilon(x)$ as described, or reduce it to an action over
one half of $S^4$ with a free boundary. 
The Euclidean Einstein-scalar action for a manifold with a boundary $B$ 
is 
\be 
{\cal S_E} = \int d^4 x \sqrt{g} \left(-{1\over 2\kappa } R +{1\over 2} 
(\nabla \phi)^2 +V(\phi)\right) - {1\over \kappa} \int_B \sqrt{h} K
\labeq{saction}
\ee
where $h$ is the determinant of the induced three-metric and 
$K$ the trace of the second fundamental form associated with the 
boundary. The last boundary term
is added to remove second derivatives from the action density
so that when we vary the action the field equations follow with 
no constraint on
derivatives of the metric normal to $B$. We 
proceed from this action, which is valid only for 
positive definite metrics, by changing coordinates on field 
space to obtain a new action 
which will be well defined even when the conformal factor vanishes. 

Under the conformal transformation, $g=\Omega^2 g^R$,
we find the Ricci scalar $R= \Omega^{-2} R^R -6 \Omega^{-3} \nabla^2  \Omega$,
$K= \Omega^{-1}K^R +3 \Omega^{-2} n^a \nabla_a \Omega$, where
$n^a$ is the unit outward normal to $B$ and
of course $\sqrt{g}= \Omega^4 \sqrt{g^R}$, $\sqrt{h}=\Omega^3\sqrt{h^R}$. 
With these substitutions and an integration by parts to remove the
second derivatives on $\Omega$  the
action (\ref{eq:saction}) becomes
\begin{eqnarray} 
\int d^4 x \sqrt{g^R} \left(-{1\over 2 \kappa } \Omega^2 R(g^R)
 -{3 \over \kappa}
\left((\nabla \Omega)^2-{\kappa\over 6} \Omega^2 (\nabla \phi)^2\right)
 +\Omega^4 V(\phi)\right)\nonumber\\
 -{1\over \kappa} \int_B \sqrt{h^R} \Omega^2 K^R.
\labeq{action}
\end{eqnarray}
The kinetic terms 
for the fields $\Omega$ and $\phi$
may be written as 
$\nabla \phi^I \nabla \phi^J G_{IJ}$ where the metric on
superspace (the space of fields) is
the matrix $G_{IJ}(\phi_K)$.  The line element 
is
therefore
proportional to
$- 6 d \Omega^2 +\Omega^2 d \phi^2$. Clearly,
$\Omega=0$ is a polar coordinate singularity of the ($\Omega$, $\phi$) 
coordinate system which may be removed by
changing to
Cartesian coordinates 
\be
\Omega_1= \Omega {\rm cosh}(\sqrt{\kappa\over 6}\phi ) \qquad 
\Omega_2=\Omega {\rm sinh}(\sqrt{\kappa\over 6}\phi),
\labeq{coords}
\ee
or light cone coordinates 
\be
\Omega_{\pm}\equiv
\Omega_1\pm \Omega_2.
\labeq{lcc}
\ee
The global Lorentzian structure
of superspace is illustrated in Figure (\ref{fig:hyper}).
The singular `point' $\Omega=0$ is now seen to actually be 
the two lines $\Omega_{+}=0$ and $\Omega_-$ =0.
We shall be interested in solutions to the field equations which
intersect these lines. 

\begin{figure}
\centerline{\psfig{file=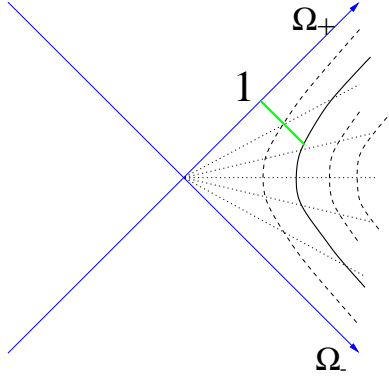,width=2.in}}
\caption{The hyperbolic geometry of superspace. The variables
$\Omega_1$ and $\Omega_2$ which are combinations of the scalar
field and conformal factor are the natural Cartesian coordinates
in which the singularities of singular instantons are resolved.
Here light cone axes are plotted, with $\Omega_\pm=\Omega_1\pm \Omega_2$.
In the conformal gauge $\Omega_+=1$ 
the solutions start at an arbitrary
positive value of $\Omega_-$ and 
ending on one branch of the light cone,
$\Omega_-=0$, as shown by the line.
}
\labfig{hyper}
\end{figure}

In these new regular coordinates the action becomes
\begin{eqnarray} 
\int d^4 x \sqrt{g^R} \left(-{1\over 2\kappa } \Omega_+\Omega_- R(g^R) -
{3\over \kappa} 
(\nabla \Omega_+\nabla \Omega_-) 
 +(\Omega_+\Omega_-)^2 
 V(\sqrt{3\over 2\kappa} {\rm ln}({\Omega_+\over \Omega_-}))\right)\nonumber\\ 
-{1\over \kappa} \int_B \sqrt{h^R} \left[K^R \Omega_+\Omega_-\right].
\labeq{actionreg}
\end{eqnarray}
This action possesses a lot of symmetry. First, there
is general coordinate invariance and conformal invariance of equation
(\ref{eq:conf}).
Second, there
is the $Z_2$ symmetry $\Omega^2 \rightarrow -
\Omega^2$ 
and 
$d^4x \rightarrow -d^4x$. To implement this symmetry 
in the regular 
coordinates we must take one of the light cone coordinates,
$\Omega_-$ say, to be odd and the other to be even.
Note that the potential $V(\phi)$ was only defined 
for real $\phi$, corresponding to positive $\Omega_-/\Omega_+$. If we
are to define the theory at negative $\Omega_-$, we must
extend the definition of the potential. The $Z_2$ symmetry
tells us how to do this, since if the action is to be
invariant under the $Z_2$, the potential must be an {\it odd} function
of $\Omega_-$. Since what enters the action is
$\Omega_-^2 V$, as long as the potential is less divergent than
$e^{+2\sqrt{2\kappa \over 3} \phi}$) as $\phi \rightarrow
\infty$ then the potential term will tend to zero and be 
only mildly nonanalytic at $\Omega_-=0$. \footnote{
There is an infinite class of potentials which are odd and for which 
$n^2 V(n)$ is analytic at 0, namely $V = a n^{-1} + b n
+c n^3...$ where $n= e^{-\sqrt{2\over 3} \phi} = \Omega_-/\Omega_+$. 
Any potential $V(\phi)$ may be arbitrarily well approximated 
by such a series over any finite range of $\phi$.}

As we approach a conformal zero, $\Omega^2 = \Omega_+\Omega_-$
tends to zero and the potential term becomes negligible. Greater 
symmetry is then revealed because the 
kinetic terms in the Lagrangian 
have an $O(1,1)$ symmetry corresponding 
Lorentz transformations on superspace, which leave 
$\Omega^2=\Omega_1^2-\Omega_2^2=\Omega_+\Omega_-$
invariant.
This symmetry does not commute with the $Z_2$ symmetry,
and in fact the combined symmetry group is Pin(1,1)
\cite{ktprep}. This symmetry is only
asymptotically exact as we approach a conformal zero,
but if we insist on preserving it we obtain important
additional constraints on the action as we now discuss.\footnote{Anomaly
cancellation is an
important motivation for the asymptotic $O(1,1)$ symmetry.
\cite{ktprep}}

As mentioned above, the presence of the orientation flip
introduces a boundary into the action being
the location of the edge of the coordinate patch 
with which we attempt to cover the entire $RP^4$. 
The presence of a boundary is an undesirable feature, since 
a boundary 
the gravitational action 
normally allows an 
arbitrary function
of the boundary geometry because the latter is not varied in determining
the equations of motion. However the situation we are discussing
is much more constrained because of the 
conformal symmetry of equation (\ref{eq:conf}), and the 
$O(1,1)$ asymptotic
symmetry which we seek to respect.
Conformal invariance 
immediately excludes terms constructed solely from 
the Riemannian metric $g^R$, such as 
the volume or the integral of the Ricci scalar. 
If we attempt to include correction factors involving
$\Omega_{\pm}$, to restore conformal invariance, 
the measure term $\sqrt{h^R}$ requires odd powers
but any curvature invariant requires even powers. Thus
we need odd powers of $\Omega_{\pm}$. But these are excluded 
by $O(1,1)$ symmetry.
Thus insisting on the symmetries of
the Lagrangian including the asymptotic $O(1,1)$,
and insisting 
the Lagrangian density be regular in the regular 
coordinates, prohibits any additional boundary 
contributions to the action apart from an irrelevant
constant.

Let us now specialise to 
$O(4)$ invariant solutions. 
The 
 Riemannian line element takes the form
\be
ds^{2 (R)} = N^2(\chi) d\chi^2 + m^2(\chi) 
d \Omega_3^2
\labeq{regmetric}
\ee 
where $N$ is the lapse function. Both $N$ and $m$ 
are arbitrary functions of $\chi\in [0,\pi)$, which is
the polar angle on the covering space
$S^4$.
The metric variables $N$ and $m$ are even parity
but $\Omega^2$ is odd. It follows that $\Omega^2$ must vanish on
the equator $\chi={\pi \over 2}$, and that the first derivatives
$m'(\chi)$ and $N'(\chi)$ must vanish there.
Since $\Omega_+$ is untwisted, and never zero, 
we may fix the conformal gauge
by setting 
$\Omega_+=1$ everywhere. In this gauge we have 
only one scalar field, namely 
\be
\Omega_- = e^{-\sqrt{2\over 3} \phi} \equiv n,
\labeq{ommm}
\ee
which by 
O(4) symmetry and oddness obeys 
$n=0$ at $\chi={\pi\over 2}$. 
Finally, when constructing the action on $RP^4$ we 
must include the function $\epsilon(x)$ encoding the orientation
reversal, and divide by two.
 O(4) symmetry forces the sign flip in $\epsilon$ to occur on the
equator, so $\epsilon=+1$ for $\chi<{\pi \over 2}$
and $-1$ for $\chi >{\pi \over2}$. We may of course
calculate the action 
by just integrating over the northern hemisphere. 
Since $\Omega^2$ is zero on the equator, the boundary
term in (\ref{eq:actionreg}) is zero. The boundary conditions
at $\chi=0$ are that the Riemannian metric and $n$  should
be regular there, so that the field equations are satisfied. 
This fixes
$N'(0)=\Omega_-'(0)=0$, 
$m(0)=0$ and  $m'(0)=1$. 

In these new variables the usual Einstein-scalar action is
\be
2 \pi^2 \int_0^{\pi\over 2}  d \chi N
 \left( 3\kappa^{-1} 
(N^{-1}(N^{-1} m')' m^2 + N^{-2} {m'}^2 m -m) n
+n^2 m^3 V(- \sqrt{3\over 2\kappa} 
{\rm ln} n)
\right).
\labeq{actiona}
\ee
It is useful to leave $N$ in the action
so that the Einstein constraint equation emerges by varying with
respect to it. But reparametrisation invariance means that 
all the field equations emerge as 
equations in the
coordinate 
$\tilde{\chi}=\int_0^{\chi}
d\chi N$, and henceforth all primes shall denote derivatives with
respect to $\tilde{\chi}$. The action may also be 
written as an integral over $0<\tilde{\chi}<\tilde{\chi}_{max}$, 
with $\tilde{\chi}_{max}$ being  the proper distance in the 
Riemannian metric from
the equator to the north pole. It is the analogue of the 
length $L$ of the circle in our non-gravitational problem.
Because we integrate over
the lapse function $N$, $\tilde{\chi}_{max}$ is  
integrated over in the gravitational path integral.

The action takes the form of that appropriate to a manifold
with a boundary, of radius $m(\tilde{\chi}_{max})$, even though
as we have emphasised there is really no boundary there.
Nevertheless there is a degree of freedom associated with
the size of the noncontractible $RP^3$ (the singular 
`point' in the Einstein frame!).
We therefore perform the path integral in two steps.
For each $RP^4$ manifold we find the noncontractible 
$RP^3$ of minimal volume. We then 
integrate over fluctuations internal to this three surface. 
Finally
 we integrate over the geometry of the $RP^3$. In the $O(4)$
invariant case, the latter is specified by
the  radius $m(\tilde{\chi}_{max})$.
The action depends on $m(\tilde{\chi}_{max})$
and for generic polynomial potentials has a minimum for one particular 
value. 

The action (\ref{eq:actiona})
yields the equations of motion,
\be 
m''n 
+{1\over 2} m'n'+{1\over 2} m n'' = -{\kappa \over 3} V(n) mn^2,
\labeq{eomsa}
\ee
\be
mn''
+3 m'n' = {2 \kappa \over 3} m n^3 {\partial V (n)\over \partial n}
\labeq{eomsb}
\ee
and the constraint which follows from varying with respect to $N$, 
\be
m'^2 n 
+m m'n' -n = -{\kappa \over 3} V(n) m^2 n^2
\labeq{eomsc}
\ee
where prime denotes derivative with respect to $\tilde{\chi} = \int N d 
\chi$. For convenience we henceforth regard the scalar potential
$V$ as a function of $n$ which as mentioned above is odd
under $n \rightarrow -n$. The equations of motion respect this
symmetry. Note that the constraint equation 
(\ref{eq:eomsc}) is consistent with the boundary conditions on the equator,
namely $n=0$, $m'=0$, as long as $n^2 V$ tends to
zero there.

\begin{figure}
\centerline{\psfig{file=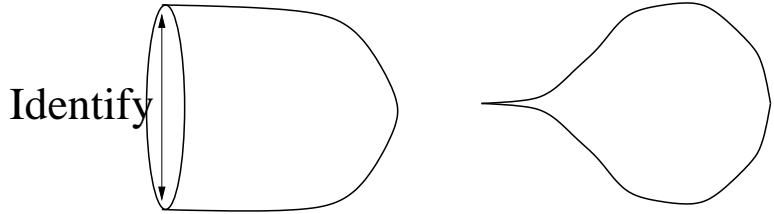,width=4.in}}
\caption{The geometry of the solutions for $RP^4$, in the Riemannian frame
(left) and in the Einstein frame (right).
}
\labfig{rp21}
\end{figure}

These field equations are merely a rewriting of those in Section 2, and
they possess the same one parameter family of 
solutions (Figure (\ref{fig:rp21})).
The boundary conditions were discussed above - at $\chi=0$ we have
$m=0$, $m'=1$ and $n'=0$, and at $\tilde{\chi}=\tilde{\chi}_{max}$ we have
$n=0$ and $m'=0$. Both $m$ and $n$ are roughly sinusoidal functions,
$m$ starting from 0 at the north pole and increasing towards a 
final value at the equator, and
$n$ decreasing from a constant at $\chi=0$ towards a linear
zero on the equator.

A particularly simple case 
is Garriga's example based on dimensionally reducing Einstein gravity
with a cosmological constant in $D=5$
\cite{garriga}, where
$V = \lambda n^{-1}$ with $\lambda$ a constant.
The equations of motion have the solutions $m=A^{-1} {\rm sin} A \tilde{\chi}$
and $n= B {\rm cos} A \tilde{\chi}$, with $A=\sqrt{\kappa \lambda/6}$
and $B$ an arbitrary constant.
The action (\ref{eq:actiona}) is zero for these solutions, 
as a result of the scale covariance of the theory.

Note that if one worked on the covering space $S^4$ and forgot
to include the $\epsilon(x)$ correction factor then one would
conclude that for any potential there is a one parameter family of regular
solutions but with zero action. These solutions would however be
half Euclidean and half anti-Euclidean. When one does
introduce $\epsilon(x)$,  
$n$ is replaced by $\epsilon(x) n$ in the action (\ref{eq:actiona}).
This is even but forced
to have zeros at the zeros of $\epsilon$ and $n$,
which are both on the equator. The field
$\epsilon(x) n$ is positive on both hemispheres but possesses
a kink on the equator. Naively, this would mean the equations
of motion were violated because the $(\epsilon n)''$ terms would introduce
a delta function on the equator. But when we vary the action 
this term arises from the variation with respect to $m$, 
and on the equator we do not vary 
$\delta m$. The point is that the minus sign needed in covariant 
derivatives of twisted fields cannot be incorporated into 
an action expressible as a single integral covering the whole
manifold. Instead we must use a constrained action.
This is analogous to the Dirac monopole case 
we discuss below where the action must be written as the sum
of two integrals, and there is a constraint on the boundary 
where the two integration regions meet.
In fact one can see in our case 
that the classical field equations, 
(\ref{eq:eomsa},\ref{eq:eomsb},\ref{eq:eomsc})
are satisfied for the entire
one parameter family of solutions, provided we 
use covariant derivatives on the twisted field $n$.
When we consider $n$ on $RP^4$, it has a linear zero on the
equatorial $RP^3$ and increases in both of the normal directions.
However the covariant derivative, as discussed in Section III,
involves introducing a relative minus sign in $n$ on either side.
Thus the covariant derivative 
is perfectly continuous in the solutions, and the field
equations are satisfied everywhere.
Satisfying the classical differential
field equations is therefore not sufficient to guarantee 
a stationary point of the 
action. In fact the action is different for the one parameter 
family of classical solutions, but there is a classical solution 
of minimal action. 

We can parametrise the solutions uniquely by the value of the
field at the regular pole $\phi_0$, or $n_0
= {\rm exp}(-\sqrt{2\kappa\over 3} \phi_0)$, analogous to 
the maximum field in Section 3. For
gently sloping potentials the equations of motion may be approximately
solved \cite{ntplb}, yielding
\be
{\cal S}_E \approx - 24 \pi^2  M_{Pl}^4
\left[ {1\over V(\phi_0)}
-{\sqrt{3\over 2} M_{Pl} V_{,\phi}(\phi_0)) \over
V^2(\phi_0)} \right]
\labeq{actval}
\ee
where $\phi_0$ is the initial scalar field value, and
$M_{Pl} = \kappa^{-{1\over 2}}$ the reduced Planck mass.
The  second 
term
is the Gibbons-Hawking surface contribution (in the Einstein frame:
as we have mentioned there is no boundary contribution in the
Riemannian frame). For
simple monomial potentials there is a minimal action solution,
typically at a value of $\phi_0$ of order $M_{Pl}$.

The minimum of the Euclidean action occurs at the 
minimal value of
$m_0=m(\tilde{\chi}_{max})$ for which a classical solution (i.e. a solution of
the equations of motion which is regular at the
north pole) occurs. To see that a minimum in $m_0$ implies 
an extremum in $S_E$, note that both are functions of the maximum
$n_0$, or the minimal field
$\phi_0$.
But if $m_0$ is minimal then $S_E$ must be stationary since
$(\partial S_E/\partial \phi_0) = (\partial S_E/\partial m_0) (\partial m_0
/\partial \phi_0) =0$. (Note that the solutions are labelled uniquely
by $n_0$ or $\phi_0$ but there are two solutions for each $m_0$).
For the one parameter family of
solutions, near 
the extremum of the action the conformal zero
is located on the minimal volume noncontractible
$RP^3$ in $RP^4$ (see Figure (\ref{fig:toby}), 
and the action $S_E$ is minimised. Thus the conformal
zero behaves rather like a brane with positive tension. 
A positive tension brane wrapped around a noncontractible $RP^3$
of minimal volume is stable on $RP^4$. This leads us to
conjecture that the minimal action Euclidean instanton in
the above construction is actually stable and has no negative modes
\cite{sgnt}.

\begin{figure}
\centerline{\psfig{file=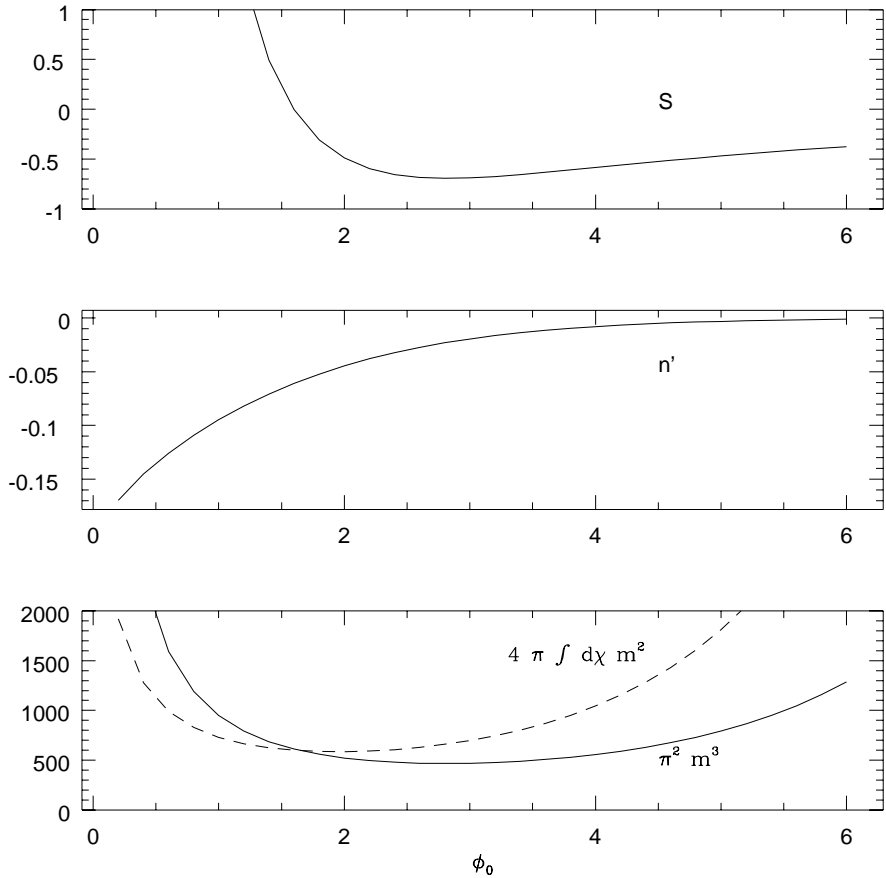,width=5.in}}
\caption{The action and other quantities describing the
the instanton solution for a simple quadratic potential $V \propto \phi^2$. 
Reduced Planck units are adopted, so that $\kappa=1$. The vertical scale
is arbitrary. The action 
possesses a minimum when the size $m$ of the noncontractible $RP^3$
is minimal. The middle graph shows the `kink' in the derivative
of the field $n$ on the conformal zero and the 
lower graph compares the volume of the $RP^3$ with that
of a three surface stretched over the north pole. As the graph shows,
the 
minimal action instanton occurs when the conformal zero is located
on the noncontractible $RP^3$ of minimal volume.}
\labfig{toby}
\end{figure}

It may seem puzzling that an action gives rise to
a one parameter family of classical solutions
of the field equations
when each solution has a different action. The resolution of
the paradox is that when the topology is nontrivial the action cannot
be expressed as a single global integral but must include constraints
present which describe the `sewing together' of the different coordinate
charts.

An analogue is provided by the Wu-Yang formulation of 
the Dirac magnetic monopole 
on $S^2$, which we briefly review. A constant radial 
magnetic field $\vec{B} = g \vec{r} /r^3$ 
of arbitrary strength 
solves the field equations $\nabla \times
\vec{B} =0$. But the energy functional
leading to ths field equation is just the integral 
over the sphere of ${1\over 2} \vec{B}^2$, and is
clearly  different for all these solutions.
The resolution as mentioned is that 
the action, which is a functional of a gauge non-invariant 
object, namely the vector potential, 
is not a single integral but has to be defined over
two coordinate patches. If we do attempt to cover the
entire $S^2$ with a single coordinate patch,  we are led to a
singular 
`Dirac string' picture. In the gravitational case above 
where we attempt to cover $RP^4$ with a single coordinate chart
we also find a `kink'
in the field $n$ leading naively to a delta function in its
equation
of motion (\ref{eq:eomsb}). 

For the magnetic monopole Wu and Yang explained how
to avoid the Dirac string.  We describe the monopole in 
two coordinate patches, 
covering the upper and lower hemisphere respectively. 
On the upper hemisphere we set the gauge potential $A_\phi=A_\phi^N= 
g(1-{\rm cos}(\theta))
$ and on the lower hemisphere $A_\phi^S=
g(-1-{\rm cos}(\theta))
$. The two are related by a gauge transformation
$A_\phi^N=A_\phi^S +\partial_\phi 
(2 g \phi)$. The total magnetic flux is given 
by Stokes theorem as $\Phi= \int_{equator} (A_\phi^N-A_\phi^S)= 4 g \pi$.
When one varies the energy functional
$\int^S + \int^N {1\over 2} \vec{B}^2$ with respect to $\vec{A}$,
the two surface terms proportional to $\int \delta 
\vec{A}\times \vec{B}$ are together proportional to the 
variation in the total magnetic flux. Thus the energy functional
is indeed stationary but only when we constrain the 
total
magnetic flux. 

In our case the analogue of the flux is the volume of the 
minimal volume noncontractible $RP^3$ on $RP^4$. This is a quantity which
is invariant under the $Z_2$ symmetry and is defined at the coordinate
overlap where the 
$Z_2$ symmetry acts, analogous to the magnetic flux above. 
For Dirac monopoles as is well known the flux is 
quantised in the presence of electrically charged fields. 
It is a natural and intriguing 
question whether there is an analogous quantisation of the
volume of the minimal $RP^3$. If there were, it would
lead to the quantisation of the density parameter $\Omega$ in the
Universe. 

Finally let us discuss the stability of these solutions under
non-$O(4)$-invariant perturbations. We shall only
give a heuristic argument in favour
of stability. There are two aspects of stability. The first
has to do with the location of the conformal zero. As mentioned,
this is analogous to a domain wall wrapped on a noncontractible
three-cycle of minimal volume, and one could expect the solution
to be stable against deformations of the wall location. 
The second relates to the freedom we had
to place the 
three-surface on which $\epsilon(x)$ switches sign on any
even parity three surface 
(Figure
(\ref{fig:coverspace}) ). 
It is straightforward to check 
the solution with least action is the one
with the zero of $\Omega_-$ located on the three-surface upon which
$\epsilon(x)$ flips sign.
This is 
seen by substituting the constraint (\ref{eq:eomsc}) 
back into the action (\ref{eq:actiona}) 
and noting that the resulting integrand is
proportional to $-6 \kappa^{-1}n +2
n^2 V$. For small $n$ the first term dominates 
and the action density is negative just above the equator,
positive below it. The most negative action is therefore
obtained by placing the $RP^3$ on the equator i.e. the 
conformal zero, so that 
the solution has non-negative conformal factor $\Omega^2$
in the entire oriented coordinate patch being considered.

\section{The Case of Two Singularities}

An interesting generalisation of the above construction is obtained
by considering the case where the underlying four dimensional manifold is
$S^{3}\times [0,1]$, with each end of the resulting cylinder completed by
a $D=4$ crosscap. The $D=4$ crosscap can be thought of as either the
total space of the twisted line bundle over $RP^3$ with fibres trimmed to
finite length, or as the space remaining when a single point is removed from
$RP^4$. Thus the topology we have in mind can be constructed by taking two
$RP^4$'s, removing a small 4-disk from each and sewing together along the
boundaries.\footnote{In fact, we can contemplate repeating this process on
any base manifold to include any
integral number $n$ of crosscaps, as is done in $D=2$ for the construction of
nonorientable Riemann surfaces. However, starting from $S^4$, only the cases
with $n\in\{0,1,2\}$ allow $O(4)$ symmetry. Taking $n=1$ gives $RP^4$, while
$n=2$ gives the second manifold under discussion.} 
The underlying Riemannian manifold 
is still described by the
metric (\ref{eq:regmetric}) but there is no regular pole, so
$m$ is positive everywhere (see Figure
(\ref{fig:kb})). The field $n$ which is $\Omega_-$ in
the gauge $\Omega_+=1$, is twisted and is zero at both the $\tilde{\chi}=0$ 
and the $\tilde{\chi}=\tilde{\chi}_{max}$ ends. As in the $RP^4$ 
example, the constraints impose $m'=0$ on the conformal zeros. 

\begin{figure}
\centerline{\psfig{file=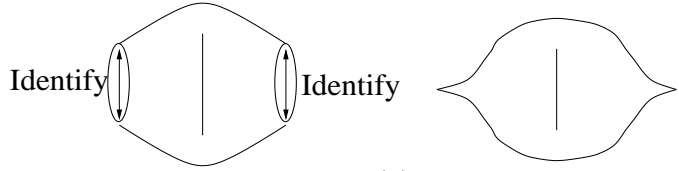,width=3.5in}}
\caption{The geometry of 
instantons which are $O(4)$ symmetric and have two singularities 
in the Einstein frame (right).
In the Riemannian frame these singularities
are blown up with a conformal transformation into two distinct non-contractible
$RP^3$'s (left).}
\labfig{kb}
\end{figure}

The solutions are parametrised by the minimal 
value of the scalar field 
$\phi_0$, or equivalently the maximum value $n_0$, 
which occurs at the midpoint $\tilde{\chi}=
\tilde{\chi}_{max}/2$. Again, with the potential 
$V=\lambda n^{-1}$, there
is a simple family of analytic solutions  namely $n= B {\rm sin} 
\sqrt{2 \kappa \lambda /3} \tilde{\chi}$, 
and $m= \sqrt{3/\kappa \lambda}$. As in the $RP^4$ example the action 
is zero for this special potential 
because of the scale covariance of the theory. 
\begin{figure}
\centerline{\psfig{file=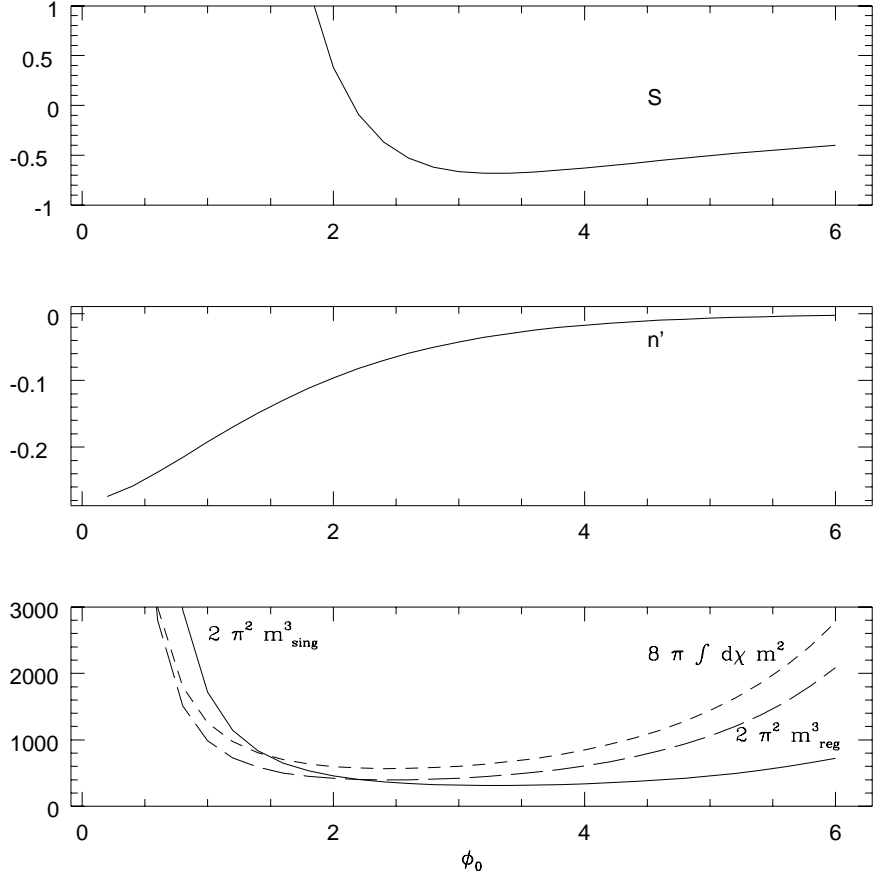,width=5.in}}
\caption{The action and other quantities describing the
the 
instanton solution with two singularities 
for a simple quadratic potential $V \propto \phi^2$.
The action 
possesses a minimum when the size $m_{sing}$ of the $RP^3$
on each end is minimal.
The middle graph shows the `kink' in the derivative
of the field $n$ on the conformal zeros. The
lower graph shows  the volume of the conformal 
zero locus in the Riemannian frame (i.e. the two $RP^3$'s),
$2 \pi^2 m_{sing}^3$ (solid line). This may be compared with
the volume of the connected three-cycle in the same homology class, 
obtained by deforming the two disconnected $RP^3$'s so that
they fuse at the centre, $8 \pi \int d \chi m^2$.
 Also shown for comparison is the 
volume of the equatorial $S^3$, $2 \pi^2 m_{reg}^3$.
}
\labfig{tobya}
\end{figure}
For a quadratic potential $V \propto \phi^2$ the action again possesses
a minimum, being approximated by the same expression
(\ref{eq:actval}) as in the $RP^4$ case.
The action and other quantities describing the
geometry are plotted in Figure (\ref{fig:tobya}).
We have checked that the `radius' $m$ increases from a minimal
value at $\pm \tilde{\chi}_{max}$ to a maximum at the centre
$\tilde{\chi}=0$. Thus as in the $RP^4$ case 
the conformal zero is located on
the non-contractible three-cycle of minimal Riemannian volume.

\section{Singular Instantons and Dimensional Reduction}

As demonstrated by Garriga \cite{garriga}, certain $D=4$ singular instanton
configurations can be realized as dimensionally reduced nonsingular
configurations of $D=5$ pure gravity with a positive cosmological constant
$\Lambda$. For concreteness, consider Euclidean gravity on a compact five
dimensional manifold $M_{5}$ with action
\begin{equation}
{\cal S}=\int_{M_{5}}d^{5}x\sqrt{g_{5}}\left[  -\frac{R_{5}}{2\kappa_5}%
+\Lambda_5\right] 
 \labeq{5daction}%
\end{equation}
The desired dimensional reduction to four dimensions amounts to compactifying
one dimension of $M_{5}$ on a circle. We write the $D=5$ line element in the
form%
\begin{equation}
ds_{5}^{2}=\exp\left[  \sqrt{\frac{2\kappa_4}{3}}\phi\right]  ds_{4}^{2}+\exp\left[
-2\sqrt{\frac{2\kappa_4}{3}}\phi\right]  \left(  \frac{d\theta}{2\pi}\right)^{2}
L^2
\labeq{dimlred}%
\end{equation}
where $ds_{4}^{2}$ is the four dimensional line element of a metric $g_{4}$ on
the dimensionally reduced space $M_{4}$, $\phi$ is a real scalar field, and
$\theta\in\lbrack0,2\pi)$ is an angular coordinate on a circle
of length $L$. The
action becomes%
\begin{equation}
{\cal S}=\int_{M_{4}}d^{4}x\sqrt{g_{4}}\left\{  -\frac{R_{4}}{2\kappa_4}+\frac
{1}{2}(\partial\phi)^{2}+\Lambda_4\exp\left[  \sqrt{\frac{2\kappa_4}{3}}\phi\right]
\right\}
\end{equation}
where $\kappa_4 =\kappa_5 L^{-1}$ and $\Lambda_4=\Lambda_5 L$.
Thus, the dimensionally reduced theory looks like gravity on $M_{4}$ plus a
minimally coupled scalar field subject to a particular potential. However, if this
scheme is to describe a locally non-singular theory in four dimensions, it
depends upon the existence of a \textit{fibration} of $M_{5}$ over $M_{4}$
with $S^{1}$ fibers. Garriga's observation is that if this topological
requirement is relaxed include certain degenerate $S^{1}$ fibrations in which fibers are allowed
to shrink to zero size, the result is a class of locally singular $D=4$
configurations which descend from perfectly regular $D=5$ configurations. In
the variables of equation (\ref{eq:dimlred}), these local singularities arise, from
the four dimensional viewpoint, as $\phi\rightarrow\infty$ poles, acompanied
by conformal zeros in the $D=4$ metric $g_{4}$. These are precisely the
Hawking-Turok singular instantons for the case of the simple exponential
potential $V(\phi)=\Lambda_4\exp\left[ \sqrt{\frac{2\kappa_4}{3}}\phi\right]  $.

Clearly, the five dimensional action of equation (\ref{eq:5daction}) has an $O(6)$
symmetric\ stationary configuration given by the round five sphere of radius
$\sqrt{\frac{6}{\kappa_5 \Lambda_5}}=\sqrt{6\over \kappa_4\Lambda_4}$. 
However, this solution does not admit a
degenerate $S^{1}$ fibration of the type described above. Instead, we pursue
an $O(4)\times O(2)$ symmetric ansatz for the metric $g_{5}$ of the form
\begin{equation}
ds_{5}^{2}=\frac{6}{\kappa_5 \Lambda_5}(d\chi^{2}+m(\chi)^{2}d\Omega_{3}^{2}+n(\chi
)^{2}d\theta^{2}) \labeq{o(4,2)ansatz}%
\end{equation}
where $d\Omega_{3}^{2}$ is locally equivalent to the $O(4)$ symmetric line
element on a unit three sphere and $\theta\in\lbrack0,2\pi)$ is again an
angular coordinate on the unit circle. This metric explicitly describes an
$S^{1}$ fibration with degenerate fibers over the locus $n(\chi)=0$. We are
led to the $D=5$ stationary configuration given by
\begin{equation}
m(\chi)=\sin(\chi)\text{ \ \ ; \ \ }n(\chi)=\cos(\chi)
\labeq{o(4,2)statconfig}%
\end{equation}
with $\chi\in\lbrack0,\frac{\pi}{2}]$.

This solution of the $D=5$ field equations looks locally like a five
sphere $S^{5}$, but as we discuss below fails \textit{globally }to give an
$S^{5}$ , describing rather the topology of the five-dimensional real
projective space $RP^{5}$. \footnote{Note that $RP^{5}$, unlike $RP^{4}$, 
is orientable.}

The desired reduction of the five-dimensional stationary configuration of
equations (\ref{eq:o(4,2)ansatz}) and (\ref{eq:o(4,2)statconfig}) 
is given by taking
\be
\phi\left(\chi\right) =-\sqrt{\frac{3}{2\kappa_4}}\log\left({2\pi\over L} 
\sqrt{\frac
{6}{\kappa_4 \Lambda_4}}\cos(\chi)\right) 
\ee
and
\be
ds_{4}^{2}={2\pi\over L} \left(\frac{6}{\kappa_4\Lambda_4}\right)^{3/2}\cos(\chi)\left(
d\chi^{2}+\sin(\chi)^{2}d\Omega_{3}^{2}\right)
\labeq{solchi}
\ee
which, from the $D=4$ point of view, clearly runs from the regular $\chi=0$ point with $\phi\left(  0\right)
=-\sqrt{\frac{3}{2\kappa_4}}\log\left(  {2\pi\over L} 
\sqrt{\frac{6}{\kappa_4\Lambda_4}}\right)  $ to the
singular $\chi=\frac{\pi}{2}$ codimension $1$ locus where the metric has a
linear conformal zero, and the scalar field $\phi$ rolls off to infinity.

We turn now to an investigation of the global topology of the five dimensional stationary
metric given by equations (\ref{eq:o(4,2)ansatz}) and
(\ref{eq:o(4,2)statconfig}). If this metric is to describe a compact five
manifold $N_{5}$, then it gives explicitly a degenerate $S^1$ fibration of
$N_{5}$ over a four dimensional submanifold $N_{4}$ of
$N_{5}$. Furthermore, the $S^1$ fibers of this fibration must shrink to
zero size precisely over a connected three dimensional submanifold $N_{3}$ of
$N_{4}$ with $N_{4}$ and $N_{3}$ locally isometric to the round $S^4$
and $S^3$, respectively. The $\chi\in\lbrack0,\frac{\pi}{2})$
subspace of $N_{5}$ which is the complement $(N_{5}-N_{3})$ is
connected. This implies via the degenerate fibration that
$(N_{4}-N_{3})$ is connected. Thus it is clear that our five dimensional
metric cannot describe a topological $S^5$, since this would imply
that $N_{4}\cong S^4$ and there can be no $N_{3}$ for which $(S^4-N_{3})$
is connected. On the other hand, the identifications $N_{5}\cong RP^5$,
$N_{4}\cong RP^4$ and $N_{3}\cong RP^3$ do work, since
$RP^n\cong S^n/Z_{2}$ and $(RP^n- RP^{n-1})$ can be connected. 

This is, in fact, the only possibility and the global topology of the
stationary five metric is that of $RP^5$, which has a degenerate $S^1$
fibration over $RP^4$, with `zero size' fibers over a
non-contractible $RP^3 \subset RP^4$. This makes explicit the connection with
our earlier analysis of Hawking-Turok singular instantons arising as
conformal zeros of four metrics wrapped on non-contractible cycles of
$RP^4$.

We can directly construct this degenerate fibration using homogeneous
coordinates on $RP^5$

\begin{equation}
(u_{0}, u_{1}, U) \sim \lambda(u_{0}, u_{1}, U), 
\forall \lambda \in (R-{0})
\labeq{homogrp5}
\end{equation}

where $u_{0},u_{1} \in R$ and
$U \in R^4$. The appropriate polar geodesic coordinate
$\chi\in\lbrack0,\frac{\pi}{2}]$ is
given by
\begin{equation}
\tan{\chi}= |U/u_{0}|
\end{equation}

For $\chi\in\lbrack0,\frac{\pi}{2})$ we can use the scaling freedom of
equation (\ref{eq:homogrp5}) to put the $RP^5$ coordinates in the form
\begin{equation}
( \xi cos{\chi}, W sin{\chi})
\labeq{s3xs1}
\end{equation}

where $\xi=u_{1}/u_{0}$ can be viewed as a local coordinate on 
$RP^1 \cong S^1$ and $W$ is a point on the unit $S^3$. The $RP^4$ base for
the degenerate fibration can then be defined in homogeneous
coordinates by $u_{1}=0$ which amounts to $\xi=0$ in the
$\chi\in\lbrack0,\frac{\pi}{2})$ subspace of $RP^5$. It only remains
to identify the submanifold of the $RP^4$ over which the fibres go to
zero size. This is the $\chi=\frac{\pi}{2}$ subspace which is
the $RP^3 \subset RP^5$ defined by $u_{0}=u_{1}=0$. Clearly, the $\chi$ of
equation (\ref{eq:s3xs1}) can be identified with the $\chi$ of the
stationary five metric given by equations (\ref{eq:o(4,2)ansatz}) and
(\ref{eq:o(4,2)statconfig}).

So Garriga's degenerate dimensional reduction emerges
as a special case of our realization of a large class of singular
instanton configurations via conformal zeros in the four metric of
Euclidean spacetime. As in our case, projective manifolds are involved.
But there are several ways in which we expect our
intrinsically four dimensional analysis to be more fundamental than an
analysis of apparently singular dimensional reductions of non-singular
$D=5$ gravitational configurations. Firstly, it is not clear why one
should insist upon starting from $D=5$. One would expect to reach
distinct and apparently singular $D=4$ configurations arising from the
reduction of theories in any $D>4$. Is the path integral to sum over
configurations in all such higher dimensions? The
intrinsic realization of singular stationary points in $D=4$ should
arise in a purely four dimensional path integral formulation. The fact
that certain of these configurations can be embedded in higher
dimensional theories may be useful but is unlikely to be
 fundamental. 
Secondly for four dimensional topologies more complicated than $S^4$
or $RP^4$, we expect Hawking-Turok type singular instantons to arise
for four dimensional conformal zeros wrapped around each distinct codimension
one integral homology cycle. It is difficult to see how these
configurations can be easily accommodated in the degenerate dimensional
reduction picture.

As a final remark, let us comment on an apparent
discrepancy between the four dimensional theory we have defined, which
has zero action for Garriga's potential, and the five dimensional
$RP^5$ theory discussed above which has a negative action corresponding to that for
a five sphere. The resolution is that in the five dimensional
interpretation the boundary condition imposed by regularity at
$\chi={\pi \over 2}$ is that $n'=1$ so that the singular
pole is regular in five dimensions. In our four dimensional
picture, however, the boundary condition is that the
Riemannian metric radius $m$ should be fixed. The actions 
appropriate to the two different boundary conditions differ
by a term which, as noted by Garriga, is minus two thirds of
the usual boundary term given in (\ref{eq:saction}). This
explains the discrepancy between the two actions. The four dimensional
action and boundary condition we have used also respects the scale
covariance $n\rightarrow \lambda n$ of the theory, which as we have 
mentioned explains why we obtain zero action. The five dimensional
boundary condition violates the scale invariance and 
that is why one obtains a nonzero action.

\section{Conclusion}

 We have discussed an interpretation of instantons 
possessing conformal zeros which links those zeros
 to the topological properties of an  underlying 
 Riemannian manifold.
 One of the interesting points to emerge
 is that the topologically twisted conformal factor does
 not admit a description via a globally defined action.
We have to use a constrained action in which
additional data is defined on an `internal boundary'. 
The classical differential equations of motion are a necessary
but not sufficient condition for the stationarity of
the resulting action, and we have shown that the 
 one parameter family 
 of singular instanton solutions discussed in Section 2
 all satisfy the equations of motion but have differing
 action. 
We have shown that in the least action solution, 
the conformal zero is located on the
minimal volume noncontractible $RP^3$ submanifold,
 and therefore behaves somewhat like a brane of positive tension.
 We generalised these arguments to singular instanton solutions 
 with two singularities which are each resolved in the
Riemannian frame into `cross-caps'possessing
 noncontractible $RP^3$ conformal zeros.
 We have  provided a regular setting for singular instantons 
 which suggests that they may be stable and possess no negative
 modes. This is an important difference with the usual
 Coleman-de Luccia instantons \cite{coldel}.

We have discussed singular instantons with one codimension one
conformal zero, which analytically continue to open
inflationary Universes. We have also discussed
instantons with two codimension one conformal zeros,
which analytically continue to closed inflationary Universes.
Examples of potentials which give realistic such Universes will
be given in a future publication.

Finally, we wish to stress that
the  idea of allowing conformal zeros is more general
and far reaching than the solutions we have explored here.
It is a first step towards discussing signature change
and topology change in four dimensional quantum geometry. 
In a forthcoming paper 
we extend the discussion  to higher codimension 
 \cite{ktprep}.

This work was supported by a PPARC (UK) rolling grant and
a PPARC studentship. We thank M. Bucher, 
S. Gratton, S.W. Hawking and V. Rubakov for helpful
comments on this work.

\end{document}